\documentclass[doublecol]{epl2}
\usepackage{amssymb,latexsym,mathrsfs}
\usepackage{amsfonts}
\usepackage{amsmath}
\usepackage{graphicx}

\newcommand{\beq}{\begin{equation}}
\newcommand{\eeq}{\end{equation}}
\newcommand{\beqn}{\begin{eqnarray}}
\newcommand{\eeqn}{\end{eqnarray}}
\newcommand{\bearr}{\begin{array}}
\newcommand{\enarr}{\end{array}}

\def\bea{\begin{eqnarray}}
\def\eea{\end{eqnarray}}
\def\ba{\begin{array}}
\def\ea{\end{array}}

\def\prl{Phys. Rev. Lett.}
\title{A  Novel Approach to Discontinuous Bond Percolation Transition}
\author{Urna Basu, Mahashweta Basu, Anasuya Kundu and P. K. Mohanty}
\institute{Theoretical Condensed Matter Physics Division, Saha Institute of Nuclear Physics,
1/AF Bidhan Nagar, Kolkata, 700064 India.}  
 
 \abstract{
We introduce a  bond percolation procedure on a $D$-dimensional  lattice    
where two neighbouring sites are  connected by $N$  channels, each operated by  valves at both ends. 
Out of a total of $N$,  randomly chosen $n$  valves are open at every site.  
A bond is said to  connect two sites if there  is at least one channel between them, which 
has open valves at both ends.  We show analytically that in all spatial dimensions, this system 
undergoes a discontinuous  percolation transition  
in the $N\to \infty$ limit when $\gamma =\frac{\ln n}{\ln N}$ crosses a threshold. 
It must be emphasized that, in contrast to the ordinary percolation models,  here the transition 
occurs even in one dimensional systems, albeit discontinuously. We also show that 
a  special  kind of discontinuous percolation occurs only in one dimension when $N$ depends on 
the system size.  
}

\begin{document}\maketitle
Percolation  transition  is  one of the most studied \cite{stauf}  critical phenomena in  non-equilibrium 
statistical physics. It is usually modeled on a lattice \cite{bond} where bonds  connecting  the 
neighbouring sites  form  independently and  randomly with probability $p$. In two and higher dimensions, 
a continuous  phase transition  to a  percolating state  having an infinitely large connected cluster occurs  when 
$p$ crosses a threshold value $p_c>0$ \cite{stauf,Grimmet}.
One dimensional systems, on the other hand, do not  show percolation transition as
these systems cannot have a percolating configuration for any 
$p<1$.  The trivial fixed point $p_c=1$, which can be approached only from the non-percolating region, 
shows signature of a  continuous phase transition.   
In this article we   propose a  generic bond 
percolation procedure which  shows a  percolation  transition, even in one dimension.  It turns out 
that  the transition is discontinuous.

Discontinuous percolation transition has been a subject of great interest since  the study of 
Achlioptas  {\it et. al.}  \cite{Achlio}.   They proposed a non-local bond percolation procedure   
which initially inhibits formation of a single large cluster.  When applied to a fully connected graph, 
it apparently results in  a  discontinuous 
transition,  which was named as {\it explosive percolation transition} (EPT). 
Different aspects of  such explosive transition  resulting from  Achlioptas  growth process 
on  several other graphs  have been studied in large number of following works\cite{o1, Fri}.
In recent studies \cite{o4} it  has been shown that  EPT can also be obtained from  percolation 
rules other than Achlioptas process. 
In fact, it has been argued that certain specific  rules \cite{Moreira} of  adding new bonds, which 
depend on the properties of the existing clusters, are sufficient  for  a system to show EPT.
However, discontinuous nature of EPT  has been questioned recently; this originated a 
long debate\cite{all} regarding the order of the transition. It has been argued that
explosive percolation is, in fact,  a continuous  transition\cite{Costa} with a small critical 
exponent $\beta$  which \textit{only} appears to be discontinuous in numerical studies.
A percolation process is proposed in the following which unambiguously shows a  discontinuous transition. 
Both the  critical point and the jump in the order parameter there  in the thermodynamic limit 
are calculated  exactly.

\begin{figure}
  \includegraphics[width=8 cm] {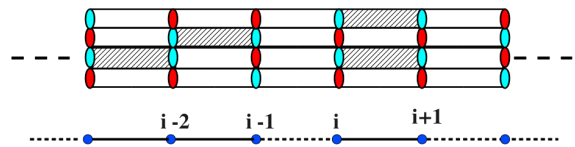}
\caption{ (Color online) The 1D model with $N=4$ channels : 
each site has $N=4$ valves (discs), of which $n=2$ are open (blue) and the rest are closed(red). 
Existence of  at least one channel which has open valves at both ends (shaded), 
represents a bond  between the corresponding sites. }
\label{fig:pic}
\end{figure}

 In this Letter we introduce a local percolation procedure on a  $D$-dimensional lattice  where 
every pair of neighbouring sites has  $N$  different  channels joining them. 
Each channel  contains a valve at every site which can either be open or closed independent 
of the other valves.  Two  sites are   connected by a  {\it bond}  if
there  exists at least one channel  which has open valves at both the sites. 
Clearly  each bond  appears randomly, independent of other bonds and  clusters.
This system shows  a discontinuous transition  in the thermodynamic limit when the relevant tuning 
parameter,  namely the  number of open valves $n$ at each site,  crosses a threshold value.  
The model could be solved exactly to locate the transition point.  
We show that, the transition is discontinuous in  all dimensions, including 1D.

First let us describe the model in  one dimension; its extension to higher dimensions is 
straight forward. The  sites labelled by $i=1,\dots,(L+1)$, are  connected to their neighbours  
by $N$ channels (see  Fig. \ref{fig:pic}). Each  channel contains  a valve at every lattice site $i$   
that can either be open or closed.  
Out of a total of $N$ valves  at every site,  $n$  are chosen randomly  and opened.  
The  neighbouring sites  of the lattice  are  said to be connected by a bond if  
there exists at least one channel between them which has open valves at both ends.
The model can be  recast  into a simpler form  by associating a set $S_i$ of $n$ integers,
randomly chosen from  a larger set $\{1,2,\dots N\}$,   to   each lattice site $i$. 
In this picture, a bond is said to connect two neighbouring sites  $i$ and $j$ if   
$S_i \cap  S_j$ is  {\it not null}.

This system, being one dimensional,  is percolating {\it only}  when  all  the $L$ 
bonds are present.  
Obviously percolation is not possible when the number of open 
valves $n$ at each site is zero. Again, the system is surely percolating for  $n>N/2$, 
as  in this  regime every pair of sites has one or more common  valves which are open. 
Our aim  here is to find  if the  system is percolating for any  non-zero $n$  smaller 
than $N/2$.

The  principal quantity of interest is the percolation probability $P_L$, which is 
defined as the probability that an arbitrary lattice site belongs to the spanning 
cluster \cite{Grimmet, stauf}. In other words, $P_L$ is the average  number of sites 
belonging to the spanning cluster and  plays the role of the order parameter
as it is nonzero  \textit{only} in the percolating regime for a 
thermodynamically large system.  For the usual bond percolation, in two and higher dimensions,
$P_L$ vanishes continuously at the critical point for $D \ge 2.$ 
It is worth mentioning that $P_L$ should not be confused with  $\Pi_L$, the probability that
there exists a spanning cluster in the system. In fact, in the thermodynamic limit
$\Pi_\infty$ for all spatial dimensions jumps from $0$ to $1$ as the connection probability 
$p$ is increased beyond $p_c.$ The fact that  $\Pi_\infty$ is  discontinuous,  is  
sometimes used \cite{stauf} to  locate the exact transition point.

Now let us calculate $\Pi_L$ for  this multi-channel model in  one dimension.
\bea \Pi_L= (1-q)^L, \label{eq:PL}\eea
where $q$ is the   probability that two neighbouring sites are not connected by 
a bond.  $\Pi_L$  vanishes  in the thermodynamic limit $L\to \infty$ 
for any $q>0$, which corresponds to the connection probability $1-q=p<1.$

For this model $q$ can be calculated as follows. 
Since $n$ valves  can be chosen from the total $N$
in $ C^N_n$ possible ways, the probability that $k$ open valves are common 
between any two neighbouring sites is 
\bea
Q_k=   \frac{  C^{N-n}_{n-k}  C^{n}_{k}}{ C^N_n }.
\eea
 Thus,  a bond between any two  neighbouring sites $i$ and $i+1$ is absent  with 
probability $q=Q_0$, which can be expressed as a function of $N$ and  
$\nu= \frac{n}{N}$ (the density of open valves at each site). Using Stirling's approximation,
\bea
q&=& {(1-\nu)\over \sqrt{1-2\nu}} \exp{[-g(\nu)N]}\label{eq:qN}\\
{\rm with~~}  
g(\nu)&=& -\ln \left[(1-\nu) \left( {1-\nu \over 1-2\nu}\right)^{(1-2\nu)}\right].\label{eq:g}
\eea 
Note that the probability $p=1-q$  of connecting  neighbouring sites by a bond, here,  
is independent of other bonds and clusters. 
 
\begin{figure}
  \includegraphics[width=8 cm] {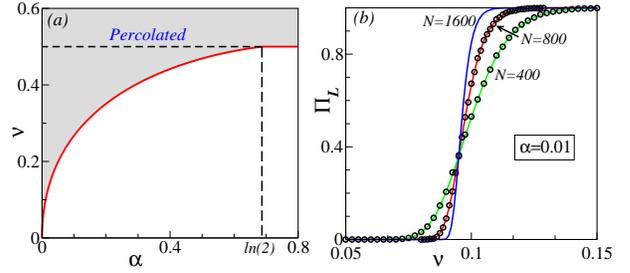}
\caption{(Color online) (a) Phase diagram in the $\alpha$-$\nu$ plane:  $\nu_c= g^{-1}(\alpha)$ is the critical line. 
For $\alpha >\ln 2$ transition occurs  trivially at $\nu_c=\frac 1 2$. 
(b) Analytical expression of $P_L$ as a function of $\nu$  (lines) for  $\alpha=0.01$ and $N=400,800~ {\rm and }~1600$  
(correspondingly  $L= e^{\alpha N}$)   are compared with  the  numerical simulations (symbols).     
The  discontinuity,  occurring at  $\nu_c= 0.095$  is more prominent  for larger $N$. }
\label{fig:phase}
\end{figure}
 From Eq. (\ref{eq:qN})  it is evident that $q$  is  finite, though exponentially small, for any given  $N$.
Thus, like other one dimensional models, here too, one cannot have a percolating state in the thermodynamic
limit $ L\to \infty$ for any arbitrary value of $n<N/2$.  However,  for  any given $L$, if one lets  
$N \to \infty$,  then $q$ vanishes, resulting in a  
percolating state for all $n>0$.   To explore the possibility of transition at a 
non-trivial  $n$, let us  couple  $N$ to the system size $L$.    
Since $q$ is an exponentially decaying function of $N$, it is suggestive that one  
takes $N= {\cal O}(\ln L)$. Let the thermodynamic  limit  be taken in  such a way  that $L$ 
approaches  $\infty$  along with $N$, whereas their ratio $\alpha = \frac{\ln L}{N}$ 
remains fixed. From  Eqs. \eqref{eq:PL} and \eqref{eq:qN}, now, 
$\Pi_L= (1- c L^{-g(\nu)/\alpha})^L$ where $c= \frac {(1-\nu)}{ \sqrt{1-2\nu}}$; 
in the thermodynamic limit, 
\bea
\lim_{L\to\infty}  \Pi_L =  \left\lbrace
 \begin{split} 
1 &~~~   {\rm for ~~} g(\nu)>\alpha \\
e^{-c} &~~~  {\rm for ~~} g(\nu) =\alpha\\
0 &~~~    {\rm for ~~} g(\nu) <\alpha.
 \end{split}\right.
\label{eq:lim}
\eea

\begin{figure}
  \includegraphics[width=8 cm] {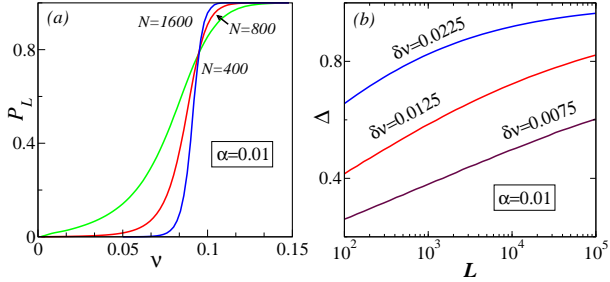}
\caption{(Color online) (a) The order parameter $P_L$ as a function of $\nu$ for $N=400,800,1600.$  
$P_L$ approches  a step function at $\nu_c=0.095$ as  $N\to\infty.$ (b) The jump in the order parameter,  
$\Delta$ (Eq. \eqref{eq:Delta}),  for different values of $\delta\nu$  approches $1$ 
as $L$ is increased.  In both the figures, the system size is $L= e^{\alpha N}$ with $\alpha=0.01.$ }
\label{fig:PL}
\end{figure}
Thus, transition from a non-percolating state ($\Pi_\infty=0$) to a percolating state 
($\Pi_\infty=1$) occurs  at $\nu_c= g^{-1} (\alpha)$.   
The corresponding phase diagram  is shown in  Fig. \ref{fig:phase}(a).  Note, that for  $\alpha > \ln 2$, 
the  transition occurs at  the trivial value $\nu_c=1/2$, which corresponds to half of the valves 
($n=N/2$)  being open at  every site.  
In Fig. \ref{fig:phase}(b)  we have  shown $\Pi_L$ versus $\nu$ 
for  three  different values of $N$ with $\alpha=0.01$. Symbols  therein  represent the  same 
obtained  from Monte-carlo simulations.  Clearly  $\Pi_L$ approaches the step function $\Theta(\nu-\nu_c)$ 
 as $L\to \infty.$
\begin{figure}
 \centering \includegraphics[width=6 cm] {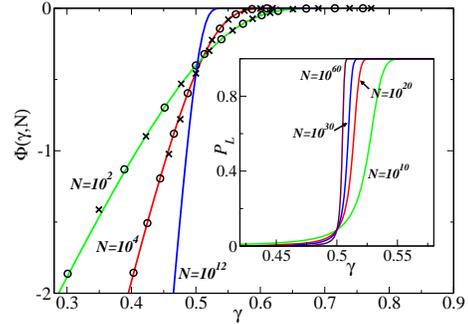}
\caption{ (Color online) $\Phi(\gamma, N)$ versus  $\gamma$ (main figure)  and  $P_L$ versus  $\gamma$  (inset)  
for $N=10^2,10^4,~{\rm and}~ 10^{12}$.  Simulation results for $\Phi(\gamma, N)$  with 
$L=4$ (circles) and $L=9$ (crosses) are also shown in the main figure. The inset shows $P_L$ for $L=100$ as 
a function of $\gamma$  for different $N.$ }
\label{fig:invLPL}
\end{figure}

In one dimension, the existence of a spanning cluster (i.e., when $\Pi_\infty=1$) implies that 
all the  sites of the lattice belong to that cluster; which  in turn  implies  that in the 
percolating regime, the order parameter  is $P_\infty=1.$  This clearly indicates that the 
percolation transition is discontinuous.  In Fig. \ref{fig:PL}(a), we have shown $P_L$ for 
$L= e^{\alpha N}$ with $\alpha = 0.01$ and $N=400,800,1600$.  To show that $P_L$,   like $\Pi_L$,  also 
approaches  $\Theta(\nu-\nu_c)$ as $L\to \infty$,  we  estimate the  jump in the order 
parameter  $P_L$ across the critical point $\nu_c$ as 
\bea
\Delta( L; \delta \nu) =  P_L (\nu_c+   \delta \nu) -P_L (\nu_c-   \delta \nu).\label{eq:Delta}
\eea
In Fig. \ref{fig:PL}(b) we plot $\Delta$ as a function 
of $L$ for $\delta\nu=0.0075,0.0125, 0.0225.$  The monotonic increase of $\Delta$  with  $L$  is 
a clear indication that  $P_\infty$  has a discontinuity at  $\nu_c.$  For any 
$\delta\nu>0$  we have  ${\displaystyle \lim_{L\to \infty}} \Delta( L; \delta \nu) =1.$
It is evident from the figure that this limit is approached  extremely  slowly.

Until now  we have been discussing   possibility of bond percolation transition in  this one dimensional  
system where both  $N$ and $L$  approach infinity keeping $ \alpha =\frac{\ln L} {N}$ fixed.   If instead 
$n$ is related to $N$, one can  obtain another percolation transition. Note, that  $q$ in Eq. \eqref{eq:qN}   
vanishes exponentially in the limit $N\to \infty$,  {\it only } when $g(\nu) > 0$.   However,  
if $g(\nu) \to 0$ slower than $1/N$, 
$q$ can approach  $1$  even in the  $N\to \infty$ limit. Let us discuss this scenario in details.

It is clear from Eq. \eqref{eq:g}   that $g(\nu)$ vanishes only at $\nu=0.$ 
To the  leading order in $\nu$ we have   $g(\nu) = \nu^2$  
and $c=1$,  thus $g(\nu)N =  n^2/N.$  
Let us take $n = N^\gamma$ with $0<\gamma <1$, so that  $\nu$ approaches  zero in the 
large $N$ limit. This results in  
\bea 
q &=& 1-p=\exp( - N^{2\gamma -1}) .   \label{eq:qgamma}
\eea

Now, in the $N\to \infty$ limit,  the  connection probability $p=1-q$ is 

\bea
\lim_{N\to\infty}  p =  \left\lbrace
 \begin{split} 
&0 &~~~&   {\rm for ~~} \gamma < 1/2  \\
&1-e^{-1} &~~~&  {\rm for ~~}     \gamma = 1/2 \\
&1 &~~~&    {\rm for ~~}   \gamma >1/2.
 \end{split}\right.
\label{eq:lim_p}
\eea 

Thus, again,  as $\gamma$ crosses the  threshold value 
$\gamma_c =\frac 1 2$, $i.e.$ when  the number of open valves at each site $n$ is increased beyond  $\sqrt N,$ 
both  $P_L$ and $\Pi_L$ in the thermodynamic limit jump discontinuously from  $0$ to $1,$ 
resulting in a discontinuous percolation transition.

It is difficult to simulate this model 
with large $L$ for the  reason that 
the limiting values of $p$  [Eq. \eqref{eq:lim_p}] 
are approached extremely slowly.  Accordingly, one needs unreasonably  large value of 
$N$ to see the transition at $\gamma_c.$ For example  one needs $N \sim {\cal O} (10^{30})$ for
a system of size  $L=100.$ 
This difficulty  can be avoided if we measure 
\bea 
\Phi(\gamma, N)= L^{-1} \ln \Pi_L= \ln(1-q)= \ln \left(1- e^{-N^{2\gamma-1}}\right) ,\label{eq:predict}
\eea 
which is independent of $L$. 
A distinct signature of  this discontinuous transition is that  the curves $\Phi(\gamma,N)$ versus $\gamma$ 
for  different values of $N$ intersect at $\gamma =\gamma_c$  as at this point $\Phi(\gamma_c,N)= 
\ln(1-e^{-1})$ is a constant. 
In Fig. \ref{fig:invLPL}  we have  plotted $\Phi(\gamma,N)$, obtained  numerically [symbols] for 
systems of size $L=4,9$ , as a function of $\gamma$  with two different  $N=10^2,10^4$
and compared  those with  Eq. \eqref{eq:predict}  [solid lines]. An excellent match  between 
these two   for different $L$s  suggests that the transition is present 
for systems of all sizes. It is only that the  required  numerical accuracy for large systems  
is hard to achieve in affordable  computational time.   A plot of  Eq.  \eqref{eq:predict} for 
$N=10^{12},$  which could  not be supplemented with the corresponding numerical results, is shown  in the same figure 
to  demonstrate  that the transition, in fact, occurs at $\gamma_c=\frac 1 2$.

Next, let us look at the order parameter $P_L$.   Since  $p(\gamma_c)= 1- 1/e$ for large $N,$
one expects that  the plots of $\Pi_L$ versus $\gamma$ for different values of  $N$ (reasonably large)  
would intersect at $\gamma_c$. This feature  is clearly visible in the inset of Fig. \ref{fig:invLPL}, 
where we have plotted $P_L$ for a system of length $L=100$ as a function of $\gamma$ 
for  $N=10^{10},10^{20},10^{30},$ and $10^{60}.$   For these large values of $N$ we have calculated 
$P_L$, the average size of the  largest cluster, by  connecting the sites with probability $p$ obtained 
directly from Eq. \eqref{eq:qgamma}. It is evident that the order parameter  is discontinuous   in the 
$N\to \infty$ limit.

\begin{figure}
  \includegraphics[width=8 cm] {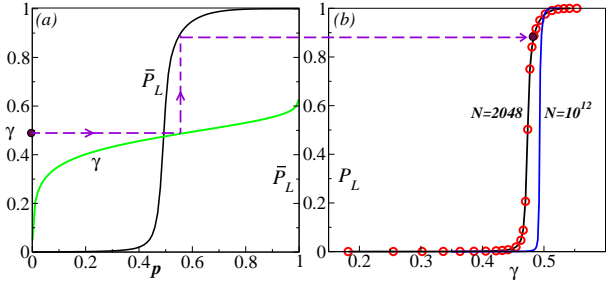}
\caption{(Color online) (a) The percolation probability  $\bar P_L$ of usual bond percolation on a $100\times100$ square 
lattice  is plotted against $p$, along with  $\gamma$  versus $p$   [calculated from  Eq. \eqref{eq:qgamma}] 
for  our model in 2D with $N=2048$. For any value of $\gamma$, which corresponds to a unique $p$ 
(thus  $\bar P_L$) one can read out $P_L$  using \eqref{eq:predict} [following the arrow].  The 
resulting $P_L$ for $N=2048$ and  $10^{12}$ are shown as solid lines in (b). Symbols therein correspond 
to $P_L$  obtained from  direct  numerical simulation  of  our model  on a $100\times100$  lattice 
with $N=2048$. }
\label{fig:barP}
\end{figure}

Let us summarize the results  obtained till now. 
We show that  a discontinuous transition  can be obtained   in this one dimensional system  
as $N$ and $L$  go to infinity, in two different ways. (A)  
Both $\alpha = \frac{\ln L}{N}$  and $\nu = \frac n N$ are  tuned to  obtain the transition, 
yielding a non-trivial phase diagram (Fig. \ref{fig:phase}(a)) in the $\alpha$-$\nu$ plane. 
(B)  $\gamma = \frac{\ln n } {\ln N}$  plays the  role of the control parameter resulting in a  phase 
transition  at $\gamma_c=\frac 1 2$.
The natural  question to ask  next is,  if any of these discontinuous transitions  is present in  higher 
dimensions.  

The model can be extended to   two dimensions in a straightforward manner; we choose to work 
on a  square lattice.  The neighbouring sites  of this   $L\times L$  lattice  
are joined  by $N$ channels. Correspondingly,  there are $N$ valves at each site which can either be open or 
closed.  An open   valve at a site allows   possibility of connection to its neighbours in {\it both}   
vertical and horizontal  directions. As before, two neighbouring 
sites of the square lattice are  connected by a bond, {\it only if} there exists  at least one channel  
which has open valves at  both these sites. Thus, the probability $q$ that  two neighbouring sites 
are  not connected by a  bond, when $n$  randomly chosen valves are opened at each site, is 
again given by Eq. \eqref{eq:qN}.

In the usual bond  percolation problem  \cite{stauf}, as mentioned earlier,  
the   percolation transition  is governed by an order parameter $P_L$ which is defined as 
the probability that a randomly selected lattice site belongs to  the spanning cluster.
The probability that a spanning cluster exists, $i.e.,$ $\Pi_L$, changes discontinuously 
across $p_c$ in the $L\to \infty$ limit, as in case of 1D.  However, unlike one dimensional systems, 
the exact formula for 
$\Pi_L$    is not known in higher dimensions.  Henceforth we concentrate only on the 
order parameter, denoted by $\bar P_L(p)$ for a $L\times L$ 
square lattice.  In the thermodynamic limit $L\to \infty,$ $\bar P_L(p)$ vanishes 
continuously at the critical threshold $p_c = 1-q_c=\frac 1 2.$

It is well known  that the  percolation probability  $\bar P_L(p)$ 
of a finite two dimensional system 
differs from  $\bar P_\infty(p)$  only by  a 
correction factor  which is negligibly small for large $L$. 
This indicates  that a discontinuous transition similar to  case (A)   
cannot be obtained in two dimension just by rendering $N$ a function of $L$.
On the other hand,    for any arbitrary $N$, the percolation probability  $P_L$ 
for this model on a $L\times L$ lattice  can be obtained from  $\bar P_L$ as
\bea
P_L  = \bar P_L (p=1-q),\label{eq:barP}
\eea
where $q$ is given by  \eqref{eq:qN}.  So, the  usual bond percolation  that occurs  
as a  continuous  phase  transition at   $p_c=1/2=q_c$  is also expected  here for any given $N$
when $\nu$  increased beyond  a  critical value  $\nu_c$  which is a  solution of $e^{g(\nu_c) N} 
= 2c(\nu_c)$.   This  continuous phase transition is similar to the 
 usual bond percolation   transition  \cite{Grimmet} on    a square lattice.

Now we turn our attention to case (B)  where $\gamma= \frac{\ln n}{\ln N }$ is used as a tuning 
parameter. Here, the  connection probability $p$  
has two distinct limiting values [see Eq. \eqref{eq:lim_p}] as  $N\to \infty$.   
So,   when $\gamma$ is  varied  continuously  in the range $[0,1]$  the connection probability $p$ jumps 
from $0$ to $1$ at $\gamma_c= \frac 1 2,$
resulting in a discontinuity in the percolation probability $P_L$ across $\gamma_c$.  Thus, 
this model shows a discontinuous transition also in  two  dimensions.

The order parameter  $P_L$  can be calculated from $\bar P_L (p)$ using Eq. \eqref{eq:barP}.
As  the analytical form of $\bar P_L (p)$  is not known,  we first  obtain the same  numerically  by simulating 
the usual bond percolation problem  on a $L\times L$  square  lattice  for $0\le p\le 1$. $P_L$ can be read 
out from this data using Eq. \eqref{eq:barP} where  $p=1-q$  is found from Eq. \eqref{eq:qgamma}  for  
any given $\gamma$  and $N$.   This procedure is  illustrated in Fig. \ref{fig:barP}(a) for a 
$100\times100$  lattice and  $N=2048$.   The  resulting $P_L$  is shown  in   Fig. \ref{fig:barP}(b)
as a solid line.    The symbols therein  correspond to the  same obtained from direct simulation of 
a $100\times100$ square lattice having   $N=2048$ channels.   It is clear from   this figure that 
the  transition occurs explosively near $\gamma_c=1/2$.  $P_L$  for $N=10^{12}$, obtained using the 
above procedure, is also shown  in Fig.  \ref{fig:barP}(b) to illustrate that, as expected, 
the transition point shifts  towards $\gamma_c=1/2$  as $N\to \infty$.

\begin{figure}
  \includegraphics[width=8 cm] {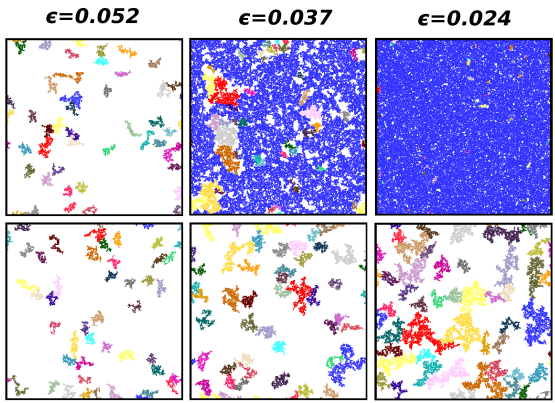}
\caption{(Color online) Typical configurations   of this model (top panel)  for 
$N=2048$ at $\gamma=\gamma_c-\epsilon$ is compared with that of  the usual bond percolation 
(bottom panel) at $p=p_c-\epsilon$. In both the cases  $50$ large clusters on a $500\times500$ 
square lattice  are shown.}
\label{fig:clust}
\end{figure}

 In contrast to the usual continuous bond percolation transition,  
the  spanning cluster appears  suddenly   in case of a discontinuous transition, as one approaches the transition point.    
In Fig. \ref{fig:clust} we have shown  the snap shots of the typical configurations ($50$ large clusters are shown) 
of  our model on a $500\times500$  lattice  near the  transition point taking $\gamma = \gamma_c -\epsilon$ and compared the 
same with the usual  bond percolation at $p= p_c -\epsilon$. 
Clearly, when $\epsilon \to 0$,  the size of the largest cluster (shown as blue) grows faster in the former 
case, indicating  the explosive nature of the transition. 

It is straight forward to  study these models on other  kinds of lattices in  two or higher dimensions.   
The explosive transition that occurs when  the number of open channels  is varied as 
$n= N^\gamma$, is quite generic.  In fact  the limiting values of  the  connection  probability 
$p$  [Eq.\eqref{eq:lim_p}] changes from being $0$ to $1$,  at $\gamma_c= \frac{1}{2}$ when  
number of channels $N\to \infty$.  Corresponding percolation probabilities change 
from $0$ to $1$  at $\gamma_c$  resulting in an explosive percolation in all dimensions. 
As discussed earlier, the other scenario where an  explosive transition  was obtained    by tuning  
$\nu= \frac n N$ and $\alpha = \frac{\ln L}{N}$, is specific to only  one dimensional systems. 

In summary,  the procedure we introduce in this article, though a simple $N$-channel extension of the usual 
bond percolation,  shows  discontinuous percolation transition in all  dimensions. 
Unlike the usual bond percolation transition, which is not possible in 1D and occurs as a continuous phase 
transition in  higher dimensions,  this model shows discontinuous transition in all $D$-dimensions, including 1D.  
The neighbours here are  connected by $N$  channels, each having an operating  valve at either 
ends. Of these  $N$ valves at each site,  $n$ are open. The neighbouring sites are  said to have a
connecting bond if  one or more  channels  joining them  are  open  at both ends. 
We show that, when $\gamma =\frac{\ln n}{\ln N}$ crosses a threshold  value $\gamma=\frac 1 2$
this  system  percolates abruptly in all spatial dimensions.  It  is rather surprising  that  a  percolation  
transition occurs even  in 1D at a non-trivial value of the tuning parameter. The  reason lies with the fact that in 
the $N\to \infty$ limit, the connection probability $p$  changes discontinuously  from $0$ to  $1$, even  though 
the tuning parameter $\gamma$  is varied continuously.

  Along with the above transition, another  explosive  percolation transition occurs especially  in   one dimension,
driven by two parameters  $\alpha = \frac{\ln L}{N}$  and $\nu= \frac nN$.   
Occurrence of this transition owes to the fact that  the percolation  probability $P_L$   in 1D explicitly depends on the system size $L$, unlike  in 
higher dimensions where system size  merely appears as a correction term. In fact  a similar transition is possible 
in the usual one dimensional bond percolation  scenario if one uses a $L$-dependent $q=L^{-z}$, $i.e.$ the 
connection probability  $p= 1-L^{-z}.$  Clearly,  in the thermodynamic limit, the percolation probability $P_L=(1-q)^L$ 
jumps from $0$ to $1$  at $z_c=1.$

We conclude with a few comments on the  differences of this model with the explosive percolation.   
  The discontinuous  change in  the size  of spanning 
cluster  have been 
reported earlier by Achlioptas {\it et. al.}  \cite{Achlio} and several following works \cite{o1,Fri,o4}  
under the name of explosive  percolation transitions.  Later studies, however,  indicate that  these transitions 
are in fact continuous \cite{Costa}, with an unusually  small critical exponent. The percolation transitions  reported here  
are  analytically proven to be  discontinuous.  
In all the studies of explosive  percolation, the connection probability is  allowed to evolve  
with the clusters, which facilitates  the sudden formation of the spanning cluster  resulting in 
an abrupt change in order parameter.  
In contrast, the connection probability of the model studied here does not at all depend on the 
existing clusters; every bond appears with the same probability.  The resulting discontinuous 
percolation is  truly an emerging behaviour.

  {\it Acknowledgements :}  UB would like to acknowledge thankfully 
the financial support of the Council of Scientific and Industrial Research, India (SPM-07/489(0034)/2007).

\end{document}